# Formation of Fractal Dendrites by Laser induced melting of Aluminum Alloys

Alexey Kucherik[1], Vlad Samyshkin [1], Evgeny Prusov [2], Anton Osipov[1,3], Alexey Panfilov [2], Dmitry Buharov [1], Sergey Arakelian [1], Igor Skrybin [1], Alexey Vitalievich Kavokin [4,5,6] and Stella Kutrovskaya [1,4,5]

1. Department of Physics and Applied Mathematics, Stoletov Vladimir State University, 600000 Gorkii street, Vladimir, Russia
2. Department of the Functional and Constructional Materials Technology, Stoletov Vladimir State University, 600000 Gorkii street, Vladimir, Russia
3. ILIT RAS Branch of FSRC Crystallography and Photonics" RAS, Shatura 140700, Russia
4. Westlake University, 18 Shilongshan Road, Hangzhou 310024, Zhejiang Province, China
5. Institute of Natural Sciences, Westlake Institute for Advanced Study, 18 Shilongshan Road, Hangzhou 310024, Zhejiang Province, China
6. Russian Quantum Center, Skolkovo IC, Bolshoy Bulvar 30, bld. 1,121205 Moscow, Russia

* Correspondence: stella.kutrovskaya@westlake.edu.cn

**Abstract:** We report on the fabrication of fractal dendrites by laser induced melting of aluminum alloys. We target boron carbide (B4C) that is one of the most effective radiation-absorbing materials which is characterised by a low coefficient of thermal expansion. Due to the high fragility of B4C crystals we were able to introduce its nanoparticles into a stabilization aluminum matrix of /1/1385.0. The high intensity laser field action led to the formation of composite dendrite structures under the effect of local surface melting. The modelling of the dendrite cluster growth confirms its fractal nature and sheds light on the pattern behavior of the resulting quasicrystal structure.

## 1. Introduction

Aluminum alloys (AAs) are promising for various industrial application due to a unique combination of the low volume density, high specific strength, corrosion resistance and thermal conductivity. High-strength aluminum alloys are significantly superior to low-carbon and low-alloy steels as well as pure titanium in terms of strength- to-density and yield-to-density ratios. According to these criteria, they approach steel alloys of higher strength and titanium alloys [1]. To further improve AAs mechanical properties the surface fabrication technologies such as ion implantation, plasma nitriding, and field assisted sintering technique (FAST) [2-4] are employed. Despite of the fact that laser processing opens up wide possibilities to locally modify the metal surface properties[5,6], aluminum and its alloys are hardly sensitive to a laser field due to the combination of high values of reflectivity, thermal conductivity and heat capacity [7]. The surrounding atmosphere plays a significant role within AAs processing that mostly leads to the formation of undesirable phases[8]. Thus, the hydrogen appearance motivates hydrated aluminum oxide growth that generally reduces the relative density of the manufactured composite matter[9]. On the other hand, the nitrogen media initiates the $Al-N$ chemical bonding which makes the whole complex quite fragile[10]. In this context, laser surface modification of AAs doped with microelements is of a considerable interest from the point of view of increasing the resulting strength and improving the temperature characteristics of the composite material. Boron carbide (B4C) is a suitable admixture candidate being an effective radiation-absorbing material that combines desirable mechanical properties even under high temperatures and a low coefficient of thermal expansion [11]. Materials of such chemical composition are characterized with a dendrite structure formed in a laser molten pool[12]. Their specific morphology is strongly dependent on the parameters of the laser action[13]. This work is aimed at the study of a laser-induced local modification of the aluminum alloy grade of AA385.0 embedded with $B_4C$ microparticles in the air atmosphere. We show that a high-energy pulse regime of laser action with an energy of up to 5 J per pulse induces the formation of two sorts of fractal clusters evolving during the crystallization process. Their morphology is highly sensitive to the thickness of a locally melted surface. The observed phenomena of the anisotropic cluster growth are in a good agreement with the results of our modelling.

## 2. Materials and Methods

Foundry of Al-Si alloy of the AA385.0 grade according to the Aluminum Association nomenclature was used as a matrix for the composite production. The powdered boron carbide F150 of 75-100 microns in particles size were used as an exogenous reinforcing component. The AAs were purchased from United company Rusal (Krasnoyarsk, Russia), the $B_4C$ powder - from Zaporozhabraziv PJSC (Zaporizhia, Ukraine) of analytical grade and used without further purification. The full fraction composition were dosed according to the calculation of the charge for the melting volume of 200 g, based on the nominal content of reinforcing particles in the composite of 5.0 wt%. Heating up to the melting point was provided by an electric resistance furnace in alundum crucibles. To increase the wettability of the $B_4C$ powder inside the melted matrix, the titanium powder of 1.0 wt% has been added before rolled to the aluminum foil. The matrix alloy was overheated up to 850°C,then the slag was removed and a suspension of the boron carbide powder was fed to the melt surface under constant stirring with a four-bladed stainless steel impeller at a speed of 300 rpm for 5 minutes. After that, the slag from the melt surface was again removed and the resulting composite suspension was poured into a vertical copper mould at a temperature of 750°C in order to produce ingots of 20 mm diameter and 100 mm height. Samples were cut off from the bottom side at a distance of 15 mm from the obtained ingots to conduct high-energy laser experiments on the surface transformation and cluster growth.

To induce the laser-alloy interaction we have used milisecond laser pulses generated by an YAG:Nd solid laser having the central wavelength of 1.06 //m, the pulse duration of 2,5ms, the repetition rate of 3.5 Hz and the variable pulse energy up to 50 J. The laser spot size was about 800 /zm, however, in our setup the laser beam was employed in a scanning regime on a rectangular area with a 1 /4 overlapping. For the detailed study of formed clusters, we have performed the scanning electron microscopy (SEM) using FEI Titan[3] with a spatial resolution of up to 2 nm and Quanta 200 3D with EDAX column with a spatial resolution of up to 7 nm.

## 3. Results

*3.1. The surface modification under laser action*

To provide surface processing of the aluminum alloy composite, the periodic set of NIR laser pulses with a fiber delivery of radiation was used. Under the effect of laser pulse action with r = 2.5ms pulse duration and the wavelength of 1.06 /zm, the temperature in the laser spot center of a *r* = 400 *ym* radius can be found from:

$$T = \frac{2q(1-R)\sqrt{\alpha\tau}}{c\sqrt{\pi}} + T_o, \qquad (1)$$

where *q* = *Ep* is a pulse energy (that has been measured experimentally and amounted either 1 J for the solid surface transformation or 5 J for the melted surface action), *R* is the aluminum reflection coefficient, *c* is the heat capacity, TQ is the starting temperature, *a.* is the thermal conductivity. In the our experiments, the temperature T was ranging from 400°C to 1000°C while the melting point of a pure aluminum is known to be 660°C. Aluminum alloys typically demonstrate a threshold behaviour [14] under laser processing. It is manifested in the reduction of the surface reflection coefficient due to the intense irradiation penetration and a vapor-gas channel formation. The specified threshold conditions mainly depend on the laser parameters such as the power density, wavelength, surface scanning speed and also on the material composition [15]. Two sorts of laser-initiated surface processing are schematically shown in Figure 1.A lightly supercritical regime in Figure 1(a) at about equilibrium conditions where a local inhomogeneity in target composition results in the appearance of a melted skin-layer. There is no any large temperature gradient this case. This is typical for diffusion processes of the skeleton-type cluster growth. On the other hand, to achieve the pronounced melting regime of a target we have used a 5J pulse energy that is highly efficient for an elongated dendrite-cluster growth, as one can see in Figure 1(b). At the ends of the dendrite branches one can see stood out B4C microcubes (see Figure 1(c)).

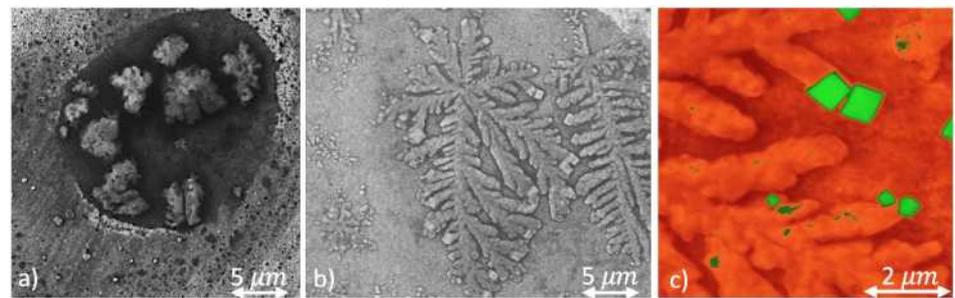

**Figure 1.** SEM images of cluster formation on an aluminum alloy surface under the effect of laser action: a) a skeleton - type cluster, formed during the local surface transformation at the pulse energy of 1J; b) a dendrite - type cluster, formed from a melted pool at the pulse energy of 5J; the panel (c) combines the SEM-image and with the results of ED AX element analysis in order to make the B4C accumulation visible (f^C-component marked by green while the alloy matrix is depicted with orange color). One can see mostly cubic lattice formation of B4C at the ends of the branches.

However, the boron carbide microparticles present in AAs play a significant role in a crystallization processes. The reinforcing particles act, on one hand, as thermal stoppers due to differences in thermal properties with the AAs matrix, and, on the other hand, as barriers to the diffusion of solute at the interface between the solid and liquid phases. During the crystallization process of the composite melted pool at relatively slow cooling that is close to equilibrium, the microparticles of the reinforcing phase are pushed by growing dendrites into the inter-dendritic regions that will be crystallized at last. The capture and displacement of crystalline elements by growing dendrite cluster is a complex phenomenon, whose nature is determined by a plenty of factors. Up to now, dozens of models were developed to describe this phenomenon, as summarised in [16]. According to the analytical Kim and Rohatgi's model [17], an outstripping of the critical crystallization velocity front would promote the capture of particles, while a velocity below the critical ones should lead to the displacement of particles by dendrite and the formation of agglomerates in the inter-dendrite regions. Thus, the value of the critical growth rate of dendrites required for a particle capture decreases with increasing size of the reinforcing particles. However, the developed models do not take to account the hydrodynamic processes that occur in the melted pool with the reinforcing particles and participate in a migration in liquid flows. Moreover, these models were originally developed for the conventional foundry and metallurgical processing. Their applicability at high local temperature gradients remains debatable. The laser action area upon melting at its energy level is mainly characterized by nonequilibrium conditions of the formation of the structure, which makes it possible to redistribute the reinforcing particles in the surface layers of the material with respect to the initial state. The velocity vector of the AAs matrix melt flow near the reinforcing particle has a decisive impact [18]. It is governed by the parameters of the laser action. In this rate, during AAs laser surface melting, the displacement of reinforcing elements can occur even at high speeds of the crystallization front movement, the speeds that significantly exceed the calculated critical values.

*3.2. Crystal growth features*

Crystals of convex rounded or faceted forms grow preserving their similarity only under a particular condition, as long as the crystal size and deviations from a counterbalance do not exceed the equilibrium values [19]. Otherwise, the crystals acquire skeleton or dendrite forms. Skeletons are the vertex and edge forms of crystals. The growth of skeletons is carried out only in the most energy-efficient directions defined by vertices and edges. Such conditions may be realized in the case of a melted media rapidly moving in the crystal growth direction. Vertices and edges directed parallel to the flow direction develop themselves much more efficiently than those that point in different directions. Dendrites appear as a result of the vertex and edge crystals growth, that occurs during an uneven diffusion of matter to the crystal. A dendrite evolves starting from each trunk of the skeleton and, as a result, the appearance of branches of the second, third and higher orders occurs. Skeletons and dendrites arise due to the unstable state of the initial convex crystal form with respect to randomly occurring perturbations. The physical cause of the appearance and development of perturbations is in the spatial anisotropy of the crystal growth speed. At the first stage, the instability of growth forms, the protrusions associated with the anisotropy of the growth speed begin outpacing other areas of the interphase boundary. The further evolution of the system into a supersaturated solution leads to the start of dendrite formation. Then each trunk of the sprawling structure grows independently to the primary protrusions and typically it is thinned in its growth to the periphery upon new lateral branches are formed. During the crystal growth process crystallizing components are supplied to the phase boundary and the thermal energy is reduced during crystallization. In the course of growth from melts, mass transfer is most often carried out with the participation of forced mixing or natural convection. In an event of growth from a melted pool, the mass transfer is mostly conducted either by forced mixing or by natural convection. Nevertheless, solid surfaces, including crystalline ones, have a fixed boundary layer, in which the transfer can be considered as occurring by ordinary diffusion or thermal conductivity. However, solid surfaces and crystalline ones have a fixed boundary layer where the transfer may be caused by a diffusion or thermal conductivity. Summarizing that, dendrites are not equilibrium products of crystal growth. They are controlled by a balance of surface tension and thermodynamic driving force at the initial stage of e.g. overcooling or oversaturation. Moreover, the anisotropy has a significant impact on the resulting shape of a dendrite, which may be further transformed by diffusion processes.

*3.3. Model of dendrite growth*

Based on the previous considerations, we formulate a model allowing to simulate the initial phase of a cluster growth. We generalise the approach formulated above by introducing the anisotropy parameter and adjusting the dimensionless latent energy, heat of transformation or anisotropy of force distributions. This generalised model allows predicting the dendrite phenomena[20,21]. The model equations operating with dimensionless variables of the phase field *p(x,y, f)* and the temperature field *T(x,y, t)* have the form:

where K is a dimensionless parameter that is proportional to the latent energy and inversely proportional to the cooling force; £ determines the melted layer thickness; £ = ar(0), where £ is the average value of £, <r(0) is an anisotropy; 6 - an angle;r is a

$$\frac{\partial T}{\partial t} = \nabla^2 T + K \frac{\partial p}{\partial t}. \qquad (2)$$

$$\tau \frac{\partial p}{\partial t} = \frac{\partial}{\partial x}\left(\varepsilon \frac{\partial \varepsilon}{\partial \theta} \frac{\partial p}{\partial y}\right) + \frac{\partial}{\partial y}\left(\varepsilon \frac{\partial \varepsilon}{\partial \theta} \frac{\partial p}{\partial x}\right) + \nabla\left(\varepsilon^2 \nabla p\right) + p(1-p)(p - 1/2 + m(T)), \qquad (3)$$

positive constant of small value; *m* sets the thermodynamic driving force as a function of tempera hire [22]. At this rate, the matter condition of *p* = 0 corresponds to the liquid phase while *p* = 1 corresponds to the solid phase. Thus, the solid-liquid phase interface is described by the last term of eq. 3. We assume the dependence m(T) in the following form: *m(T)* = tanh/T-T)ir where a is a positive constant (a < 1). |m(T)| < 1/2 for the considered temperature range. To describe the anisotropy, the function <r(0) is taken as <r(0) = 1 + *3*cos *j(f)* − 0Q), where is an angle at the maximum value of £. The parameter *3* defines the strength of anisotropy and *j* is the mode number of the anisotropy [23]. The model equations were discretized on the computational domain of a uniform grid. To solve the equations, a simple explicit scheme on a 4-point template has been used[24]. The model variational parameters are: dimensionless latent energy *K,* the anisotropy strength *3* and the anisotropy mode number *j*. Assuming these dimensionless constants to be given by £ = 10$^{-2}$; т = 3 * 10$^{-4}$; *a* = 0.9; 7 = 10; $T_e$ = 1, we have used a stable explicit scheme to perform the calculations. The calculation results and their discussion will be given in the next section. Our approach correctly describes the growth regime where the newly formed admixture crystallites start playing an important role in the anisotropy of the growth process. Thus, to take into accoimt the cubic $B_4C$ lattices' occurrence and its effect acting on a embryo's shape, in the second stage we modified the calculation algorithm with the diffusion limited agglomeration (DLA) elements. Here, the further cluster growth was modeled assuming the cluster element migration from the boimdaries (schematically illustrated in Figure2), while the calculation area was divided into subdomains: the area of cluster's origin $R_p$, where a new cluster element comes from; the outer area $R_e$, that was located enough far from the cluster, and its role was on the element annihilation while it was getting output; the minimum coverage area of the cluster R;,, which covers the entire formed aggregate [25].

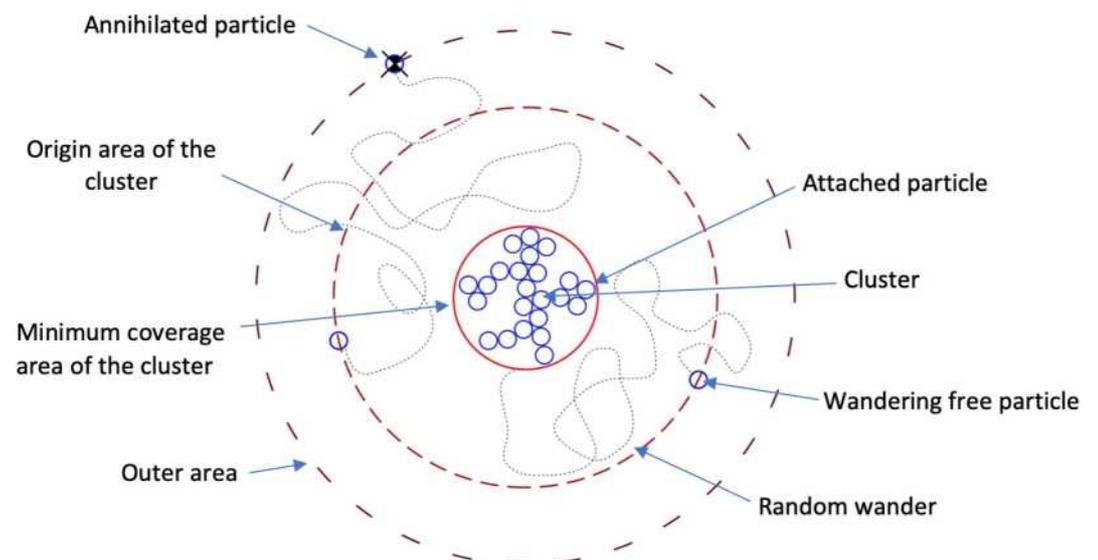

**Figure 2.** The schematic concept of the DLA growth algorithm with a minimum specified coverage area .

The adhesion probability (s) is a factor determining the dynamics of cluster growth modeling with DLA approach where it represents the relative diffusion coefficient and takes values from S = (0; 1]. So, the selection of the embryo shape at the initial stage of cluster growth, and then the adhesive consideration also allows us to adequately describe the spatial features of the experimentally obtained clusters.

## 4. Results and Discussion

We localised the embryo in the center of the calculation area. Then varying the K-parameter, we obtained critical embryos of different form-factors (see the embryos on Figure 3 (a-b) and f-g)). In the experiment, it depends on the temperature of the central region of laser-target interaction and, as a consequence, depends on a thickness of the melted layer. In Figure 3 one can see the calculation results obtained for 2 different sorts of laser surface processing: local surface transformation (see the top panel of Figure 3) and cluster crystallization from a melted pool (see the bottom panel of Figure 3).

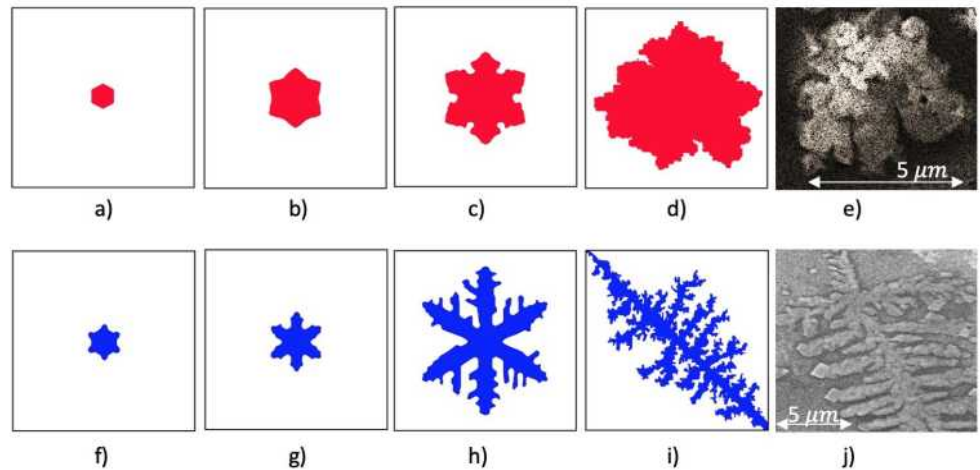

**Figure 3.** The calculation results and their comparison with real detected clusters: the top panel shows the dendrite-snowflake generation at K = 1.2; $\delta$ = 0.050; $/$ = 6; $and0$ = $1/2\pi$, S = 1: t=100 (a), t=300 (b), t=1000 (c), t=3500 (d) and corresponding to this model growth, SEM image of a snowflake cluster formation on (e); the bottom panel shows dendrite-star generation at $K$ = 1.6; J = 0.040; $/$ = 6; $and0$ = $\sqrt{2}n, S$ = 1: t=100 (f), t=300 (g), t=1000 (h), t=3500 (i) and corresponding to this model growth, SEM image of a star cluster on (j).

To compare the simulated clusters with the experimental ones, the fractal dimensions were estimated using the method of boxcounting [26]. The fractal dimensions $D$, certify of a good agreement between calculation and experimental data with an accuracy of 5% for dendrite-snowflake in Figure 3(d) and (e) with $D$ = 1.91 and for dendrite-star in Figure 3(i) and (j) with D = 1.57.

## 5. Conclusions

In conclusion, we have observed two types of a clusters growth on a AAs surface under different regimes of a laser processing. Under the effect of melted matter of a aluminum matrix composite target the dendrite clusters framed with B4C microcubes have been formed. The structure of detected clusters have successfully been modelled with the approach of initial phase of a cluster growth modified with a particle's migration effect. This results open a way to fabricate a new class of alloys sculptured with refractory boron carbide microelements.

**Author Contributions:** AK conceived the work; VS performed the laser setup; EP and AP contributed to samples preparation; AO contributed to the laser ablation experiments; DB contributed to cluster growth modelling, SA performed fractal dimension calculations and contributed to the interpretation; IS realized SEM study; AVK contributed to the original draft preparation and manuscript revision; SK contributed to the interpretation of the result and wrote the manuscript. All authors have read and agreed to the published version of the manuscript.

**Funding:** The work of SK and AVK is supported by the Westlake University, project No. 041020100118 and the Program 2018R01002 funded by Leading Innovative and Entrepreneur Team Introduction Program of Zhejiang. This work is also partially supported by RFBR grants 19-32-50095, 19-32-90085, 20-21-00038. A.O. acknowledges the support from Ministry of Science and Higher Education within the State assignment Vladimir State University, project No. 0635-2020-0013.

**Institutional Review Board Statement:** Not applicable.

**Informed Consent Statement:** Not applicable.

**Conflicts of Interest:** The authors declare no conflict of interest.

## Abbreviations

The following abbreviations are used in this manuscript:

SEM     Scanning Electron Microscopy
NPs      Nanoparticles
HSLA High-strength low-alloy steel
FAST Field assisted sintering technique
EDAX Energy-dispersive X-ray analyser
DLA     Diffusion limited agglomeration
NIR      Near infrared spectral range